\def\bm#1{{\mbox{\boldmath$#1$\unboldmath}}}

\def\dert#1#2{\frac{{{d}}{#1}}{{{d}}{#2}}}              

\def\bar{\begin{eqnarray}}
\def\ear{\end{eqnarray}}

\def\eqI{\begin{equation}}
\def\eqF{\end{equation}}
\def\eqIa{\begin{eqnarray}}
\def\eqFa{\end{eqnarray}}
\def\rp#1#2{{#1\over#2}}

\def\lb#1{\label{#1}}
\def\beq{\begin{equation}}
\def\eeq{\end{equation}}

\def\oc2{$\mathcal{O}(c^{-2})$}

\documentclass[nofootinbib]{revtex4}

\usepackage{amsmath,amsthm,amscd,amssymb}
\usepackage{latexsym}
\usepackage{graphicx,epsfig}

\begin{document}

\title{Gravitomagnetic time-varying effects on the motion of a test particle}

\author{Matteo Luca Ruggiero}
\email{matteo.ruggiero@polito.it}
 \affiliation{UTIU, Universit\`a Telematica Internazionale Uninettuno, Corso Vittorio Emanuele II 39, 00186 - Roma, Italy}%
 \affiliation{Dipartimento di Fisica, Politecnico di Torino, Corso Duca degli Abruzzi 23, 10129 - Torino, Italy}

\author{Lorenzo Iorio}
\email{lorenzo.iorio@libero.it}
\affiliation{INFN-Sezione di Pisa. Address for correspondence: Viale Unit$\grave{a}$ di Italia 68
70125 Bari (BA), Italy.}

\date{\today}

\begin{abstract}
We study the effects of a time-varying gravitomagnetic field on the motion of test particles. Starting from recent results, we consider the gravitomagnetic field of a source whose spin angular momentum has a linearly time-varying magnitude. The acceleration due to such a time-varying gravitomagnetic field is considered as a perturbation of the Newtonian motion, and we explicitly evaluate the effects of this perturbation on the Keplerian elements of a closed orbit. The theoretical predictions are compared with actual astronomical and astrophysical scenarios, both in the solar system and in binary pulsars systems, in order to evaluate the impact of these effects on real systems.
\end{abstract}


\maketitle


\section{Introduction} \label{sec:intro}

The similarity between Newton's law of gravitation and Coulomb's law of electricity has been largely investigated since
the nineteenth century, and many authors suggested that, in analogy with the electromagnetic case, the motion of masses
could produce a \textit{gravitomagnetic} (GM) field \cite{ruggiero02}. Actually, in General Relativity (GR) a GM field is indeed generated by the mass current: more in general, any theory that combines Newtonian gravity with Lorentz invariance must include a GM field.
It is well known that the field equations of GR, in linear post-newtonian approximation, can be written in terms of Maxwell equations for the gravitoelectromagnetic (GEM) fields \cite{mashhoon01}, \cite{mashhoon03} (for some limitations of this analogy see \cite{ps}, while for a different approach to  gravitoelectromagnetism based on the use of tidal tensors in GR, see \cite{p1,p2}). In this context, the role of the gravitational induction has been recently investigated \cite{bini08}, and it has been shown that it is possible to describe gravitational induction in analogy with the Faraday-Lenz law of electromagnetism. Furthermore, for stationary GEM fields, the spatial components of the geodesic equation reduce, up to first order in $v/c$, to the Lorentz force law, while for general time-varying fields the corresponding expression is more complicated. Time-varying GEM fields can be interesting in actual astrophysical events: for instance when a star, or a planet, loses its spin angular momentum, it generates a time-varying GM field, whose effects on the neighboring objects can be evaluated. As an example, one may consider the decrease of the rotation rate of the Earth (due to tidal effects): this corresponds to a loss of  angular momentum which, in turn, produces a time-dependent GM field; however, this variability is so small that is not expected to influence experiments aimed at the measurement of the GM field of the Earth, such as GP-B \cite{mashhoon08,gpb}. 

In this paper we are concerned with the study of the effects of time-variability of the GM field on test particles. Starting from the results obtained in \cite{bini08}, we consider a non-stationary spacetime where the magnitude of the spin angular momentum of the source of the gravitational field varies linearly with time (the case of the variation of the direction of the spin angular momentum for the effects of external torques was considered in \cite{Iorio2002}, while the instability  of closed Keplerian orbits induced by a time-varying 
gravitomagnetic field was studied in \cite{mashhoon08a}). In particular, we focus on the equation of motion of a test particle, where we consider the relativistic linear GM contribution as a perturbation  of the Newtonian motion.  The effects of such a perturbation on the Keplerian elements of a closed orbit are evaluated, and then compared with some actual astronomical and astrophysical events,  in order to estimate the impact of these effects on real systems.

\section{Motion in a non-stationary gravito-magnetic spacetime} \label{sec:motion}

The solutions of the GR field equations, for a localized slowly rotating source, in linear approximation give raise to the following spacetime metric (see \cite{mashhoon03})
\beq
\label{eq:metric}
d s^2= -c^2 \left(1-2\frac{\Phi}{c^2}\right)\rm d t^2 -\frac4c ({\mathbf A}\cdot \rm d {\mathbf x})\rm d t +
 \left(1+2\frac{\Phi}{c^2}\right)\delta_{ij}\rm d x^i \rm d x^j\ ,
\eeq
where the GEM potentials can be expressed as
\beq
\label{eq:potentials0}
\Phi=\frac{GM}{r}, \quad {\mathbf A}=\frac{G}{c}\frac{{\mathbf S}\times {\mathbf r}}{r^3}\ ,
\eeq
in terms of the mass $M$ and the spin angular momentum ${\mathbf S}$ of the source; in what follows the Cartesian coordinates $\{x,y,z\}$ are choosen in such a way that the angular momentum is parallel to the axis z: $\hat {\mathbf S} \parallel {\bm e}_{z}$, where $\hat {\mathbf S}$ is the unit vector in the direction of ${\mathbf S}$; furthermore, we use the notation $|\mathbf{r}|=r$. In order to solve the field equations the transverse gauge has been imposed, which, in terms of the GEM potentials reads
\beq
\frac{\partial \Phi}{\partial t}+\nabla \cdot \left(\frac 1 2 \mathbf{A} \right)=0. \label{eq:gauget}
\eeq

In particular, in \cite{bini08}  GEM potentials in the form
\beq
\label{eq:potentials}
\Phi=\frac{GM}{r}, \quad {\mathbf A}=\frac{G}{c}\left[S_0+S_1\left(t-t_{0} \right) \right]\frac{\hat {\mathbf S}\times {\mathbf r}}{r^3}\ ,
\eeq
have been considered, where the magnitude of the angular momentum varies linearly with time\footnote{In order to carry out the perturbative analysis of the motion, we have chosen the constant $S_{0}$ in such a way that it corresponds to the magnitude of the angular moment at $t=t_{0}$, i.e. when the particle is at the periastron of its unperturbed  orbit (see below).} $S=S(t-t_{0})\doteq \left[S_0+S_1\left(t-t_{0} \right)\right]$, and it has been shown that the metric (\ref{eq:metric}), with these potentials, represents a solution of the linearized GR field equations, in the linear GEM approach, in which all terms of $O(c^{-4})$ in the metric tensor are neglected. Of course, this time-varying solution is consistent with the linear GEM approach if the time variation of the magnitude of the angular momentum does  not  break the validity of the linear approximation\footnote{In other words, we suppose to work within a time interval $\Delta t$ for which $2|\mathbf{A}| \ll c^{2}$, so that the gravitomagnetic term (i) exists and (ii) can be considered as a perturbation of the flat spacetime metric.}; moreover, the instants of time at which the temporal variation starts and ends are not considered. 

Actually, it is interesting to point out that the effective source for the spacetime metric (\ref{eq:metric}) with the GEM potentials in the form (\ref{eq:potentials}) is localized and divergenceless, therefore its volume integral vanishes, even though stresses  distributed throughout space arise, which fall off as $r^{-3}$ for $r \rightarrow \infty$: however this fact does not affect the viability of the linear GEM approach, since the effects of these stresses on the metric tensor are of $O(c^{-4})$ (see the discussion in \cite{mashhoon08} and \cite{bini08}).\\

The spatial components of the geodesic equation of the metric (\ref{eq:metric}), with the GEM potentials (\ref{eq:potentials}) up to the dominant gravitomagnetic terms are (see \cite{bini08})
\begin{eqnarray}
\label{eq:geo2}
\frac{ d{\mathbf v}}{ dt}+\frac{GM{\mathbf r}}{r^3}&=& \frac{2G}{c^2}\frac{\dot {\mathbf S}\times {\mathbf r}}{r^3}-\frac{2}{c}{\mathbf v}\times {\mathbf B},\
\end{eqnarray}
where the gravito-magnetic field is
\beq
{\mathbf B}= \frac{G}{c}(S_0+S_1t)\frac{1}{r^3}[3\,(\hat {\mathbf S}\cdot \hat {\mathbf r})\hat {\mathbf r}-\hat {\mathbf S}]\label{eq:defB11}
\eeq
The geodesic equation (\ref{eq:geo2}) can be written in the form
\begin{eqnarray}
\label{eq:geo3}
\frac{d {\mathbf v}}{d t}+\frac{GM{\mathbf r}}{r^3}&=& \mathbf{W},
\end{eqnarray}
where $\mathbf{W}$ can be thought of as an acceleration perturbing the Newtonian motion.  We set
\beq
\mathbf{W}=\mathbf{f}_{1}+\mathbf{f}_{2},  \label{eq:defF1F2}
\eeq
where
\beq
\mathbf{f}_{1}=\frac{2G}{c^2}\frac{\dot {\mathbf S}\times {\mathbf r}}{r^3},
\label{eq:defF1}
\eeq
and
\beq
\mathbf{f}_{2}=-\frac{2}{c}{\mathbf v}\times {\mathbf B}. \label{eq:defF2}
\eeq
The latter term is nothing but the gravitomagnetic Lorentz acceleration, while the former is an azimuthal acceleration due to the time variability of the angular momentum $\mathbf{S}$.

\section{Perturbation of the Newtonian motion} \label{sec:pert}

In order to calculate the impact of the acceleration (\ref{eq:defF1F2}) on the orbit of a test particle, let us project it onto the radial ($\bm{e}_{r}$), transverse ($\bm{e}_{t}$) and normal ($\bm{e}_{n}$) directions of the co-moving frame picked out by the three unit vectors. To this end let us briefly describe  the geometry of the problem we are dealing with: we consider a first set of Cartesian coordinates $\{x,y,z\}$, whose origin coincides with the source of the gravitational field; in particular, in this coordinates we have $\hat {\mathbf S} \parallel {\bm e}_{z}$. Then, we introduce another set of Cartesian coordinates $\{X,Y,Z\}$, with the same origin: the $X$ axis is directed along the ascending node, the $Z$ axis is perpendicular to the orbital plane. The angle between the $x$ and $X$ axis is $\Omega$, the longitude of the ascending node, while the angle between the $z$ and $Z$ axis is $I$, the inclination of the orbital plane.  To summarize, we write the relations between the basis vectors

\begin{eqnarray}
\bm{e}_{X} & = & \cos \Omega \ \bm{e}_{x} + \sin \Omega \ \bm{e}_{y}, \nonumber \\
\bm{e}_{Y} & = & -\cos I \sin \Omega \ \bm{e}_{x} + \cos I \cos \Omega \ \bm{e}_{y} + \sin I \ \bm{e}_{z}, \label{eq:tras1} \\
\bm{e}_{Z} & = &  \sin I \sin \Omega \ \bm{e}_{x} - \sin I \cos \Omega \ \bm{e}_{y} + \cos I \ \bm{e}_{z}. \nonumber
\end{eqnarray}

Let ${\mathbf{r}}=r (f) \bm{e}_{r}$ be the
Keplerian orbit. The radial unit vector $\bm{e}_{r}$ in the plane of motion  is described by
\begin{equation}
\bm{e}_{r} = \cos u
\ \bm{e}_{X}+\sin u \ \bm{e}_{Y}, \label{eq:orbita1}
\end{equation}
where $u = \omega+f$,   $\omega$ is the argument of periastron  and $f$ is the true
anomaly. We remember that the unperturbed closed orbit, i.e. the orbit under the  Newtonian acceleration in eq. (\ref{eq:geo3}), is an ellipse described by
\begin{equation}
r (f)= \frac{a(1-e^2)}{1+e\cos f}, \label{eq:kep1}
\end{equation}
where $a$ is the  semi-major axis and $e$
is the eccentricity.

We may write the velocity in the form
\beq
\mathbf{v}=\frac{na}{\sqrt{1-e^{2}}} \left[ \left(-\sin u-e \sin \omega \right)\bm{e}_{X}+\left(e\cos \omega+\cos u \right)\bm{e}_{Y}  \right], \label{eq:defv1}
\eeq
where  $n=\sqrt{GM/a^3}$ is the un-perturbed Keplerian mean motion related to the un-perturbed Keplerian orbital period by $P_{\rm b}=2\pi/n$.
Now, on using eqs. (\ref{eq:orbita1})--(\ref{eq:defv1}) we can calculate the expression of the perturbing accelerations (\ref{eq:defF1}),(\ref{eq:defF2}) along the unperturbed orbit. In particular,  we are interested on the components of these accelerations along the radial, tangential and normal (to the plane) directions, which, if we choose without loss of generality, $\Omega=0$, are defined by the following unit vectors
\begin{eqnarray}\
\bm{e}_{r} &=& \cos u\ \bm{e}_{x} + \cos I \sin u\ \bm{e}_{y} + \sin I \sin u\ \bm{e}_{z}, \label{eq:deferxyz} \\
\bm{e}_{t} &=& -\sin u\ \bm{e}_{x} + \cos I \cos u\ \bm{e}_{y} + \sin I \cos u\ \bm{e}_{z} \label{eq:defetxyz} \\
\bm{e}_{n} &=& -\sin I\ \bm{e}_{y} + \cos I \ \bm{e}_{z}, \label{eq:defenxyz}
\end{eqnarray}
In order to take into account the explicit time dependence, it is useful to introduce the eccentric  anomaly $E$, which is related to the
true anomaly $f$ by the following relations:
\beq \cos f = \rp{\cos E - e}{1-e\cos E},\quad \quad \sin f = \rp{\sqrt{1-e^2}\sin E}{1-e\cos E} \label{eq:csphiE}.\eeq
Furthermore, in terms of the eccentric anomaly we may write the unperturbed ellipse in the form
\beq r = a(1-e\cos E),\label{eq:keprE}\eeq
the time dependence can be expressed in terms of the eccentric anomaly thanks to
\beq t-t_0 = \rp{E-e\sin E}{n},\label{eq:tE}\eeq
where $t_{0}$ corresponds to the passage at the periastron.

That being said, we are now able to explicitly calculate the components of the perturbing acceleration $\mathbf{W}$, in the form
\begin{eqnarray}
W_r&\doteq&\left(\mathbf{f}_{1}+\mathbf{f}_{2}\right) \cdot \bm{e}_{r}, \nonumber \\
W_t&\doteq&\left(\mathbf{f}_{1}+\mathbf{f}_{2}\right) \cdot \bm{e}_{t}, \label{eq:acc1}\\
W_n&\doteq&\left(\mathbf{f}_{1}+\mathbf{f}_{2}\right) \cdot \bm{e}_{n},  \nonumber
 \end{eqnarray}
and we obtain
\begin{eqnarray}
W_r & = & \frac{2G\cos I \sqrt {1-{e}^{2}}}{c^{2}a^{2}}\,{\frac {n S(E)}{\left[1-e\cos E \right]^{4} } },
 \label{eq:accwr} \\
  W_t & = & \frac{2G\cos I}{c^{2}a^{2}}  \frac {S_{1}-nS(E) e \sin E-2S_{1}e \cos E+S_{1}e^{2}\cos^{2} E}{\left[1-e\cos E\right]^{4}},
  \label{eq:accwt}  \\
W_n & = &  \frac{2G\sin I}{c^{2}a^{2}\sqrt{1-e^{2}}}S_{1}\left\{ \frac{\sin \omega \sin  E \left(1-e^{2} \right) + \cos  \omega \sqrt{1-e^{2}} \left[ e-\cos E \right]  }{\left[1-e\cos E\right]^{3}} \right\} + \label{eq:accwn} \\
          & + & \frac{2G\sin I}{c^{2}a^{2}\sqrt{1-e^{2}}}nS(E)\left\{ \frac{2\sin  \omega  \cos  E+2 \cos \omega \sin  E\sqrt{1-e^{2}}}{\left[1-e\cos E \right]^{5}} \right\} + \nonumber \\
          & - & \frac{2G\sin I}{c^{2}a^{2}\sqrt{1-e^{2}}}nS(E)\left\{ \frac{3 \sin  \omega -\sin  \omega  \cos^{2} E -\cos  \omega \sin  E \cos  E	 \sqrt{1-e^{2}}}{\left[1-e\cos E \right]^{5}} \right\}e + \nonumber \\
          & - & \frac{2G\sin I}{c^{2}a^{2}\sqrt{1-e^{2}}}nS(E)\left\{ \frac{3 \cos  \omega \sin E \sqrt{1-e^{2}}+2\sin \omega \cos E}{\left[1-e\cos E\right]^{5}} \right\}e^{2} + \nonumber \\
          & - & \frac{2G\sin I}{c^{2}a^{2}\sqrt{1-e^{2}}}nS(E)\left\{ \frac{\sin \omega \cos^{2} E-3 \sin  \omega}{\left[1-e\cos E\right]^{5}} \right\}e^{3}, \nonumber
\end{eqnarray}
where
\beq
S\left(E \right) = S_{0}+\frac{S_{1}E}{n}-\frac{S_{1}e \sin E}{n}. \label{eq:defSE}
\eeq

The components (\ref{eq:accwr}-\ref{eq:accwn}) of the perturbing acceleration must be inserted  into the right-hand-side of the Gauss equations  for the variations of the Keplerian orbital elements:
\begin{eqnarray}\lb{Gauss}
\dert a t & = & \rp{2}{n\sqrt{1-e^2}} \left[e W_r\sin f +W_t\left(\rp{p}{r}\right)\right],\lb{gaus_a}\\
\dert e t  & = & \rp{\sqrt{1-e^2}}{na}\left\{W_r\sin f + W_t\left[\cos f + \rp{1}{e}\left(1 - \rp{r}{a}\right)\right]\right\},\lb{gaus_e}\\
\dert I t & = & \rp{1}{na\sqrt{1-e^2}}W_n\left(\rp{r}{a}\right)\cos (f+\omega),\\
\dert\Omega t & = & \rp{1}{na\sin I\sqrt{1-e^2}}W_n\left(\rp{r}{a}\right)\sin (f+\omega),\lb{gaus_O}\\
\dert\omega t & = &\rp{\sqrt{1-e^2}}{nae}\left[-W_r\cos f + W_t\left(1+\rp{r}{p}\right)\sin f\right]-\cos I\dert\Omega t,\lb{gaus_o}\\
\dert {\mathcal{M}} t & = & n - \rp{2}{na} W_r\left(\rp{r}{a}\right) -\sqrt{1-e^2}\left(\dert\omega t + \cos I \dert\Omega t\right),\lb{gaus_M}
\end{eqnarray}
where ${\mathcal{M}}$  is mean anomaly of the orbit of the test particle and  $p=a(1-e^2)$ is the semi-latus rectum.  In order to obtain the secular effects induced by the perturbing acceleration  we must evaluate the Gauss equations onto the unperturbed Keplerian ellipse (\ref{eq:keprE}), taking into account eqs. (\ref{eq:csphiE}) and average\footnote{We used $\dot f+\dot \omega=\dot f$ because over one orbital revolution the pericentre $\omega$ can be assumed constant.} them over one orbital period $P_{\rm b}$ of the test particle  by means of
\beq \rp{dt}{P_{\rm b}} = \left(\rp{1-e\cos E}{2\pi}\right)dE.\lb{kep_dt}\eeq
Actually we are interested in the explicit calculation of the secular variations for orbits that have small eccentricities, as a consequence  we neglect terms of order $\mathcal{O}(e^2)$. What we get is:
\begin{eqnarray}\lb{orbi}
  \left\langle\dot a\right\rangle &=& 4 \frac{GS_{1}}{c^{2}a^{2}n}\cos I, \label{eq:vara}\\
  \left\langle\dot e\right\rangle &=& -\frac{GS_{1}\left(2+e \right)}{c^{2}a^{3}n} \cos I, \label{eq:vare} \\
  \left\langle\dot i\right\rangle &=&  -2 \frac{G S_{1} \left(1+e \right)\cos^{2}\omega}{c^{2}a^{3}n}\sin I \label{eq:vari} \\
  \left\langle\dot\Omega\right\rangle &=& 2 \frac{G}{c^{2}a^{3}n}\left\{-S_{1} \left[\frac 1 2 \sin 2 \omega\left(1+e \right)-\pi\right]+S_{0}n \right\} \label{eq:varOmega} \\
  \left\langle\dot\omega\right\rangle &=& - 2\frac{G}{c^{2}a^{3}n}\left\{-S_{1}\left[\frac 1 2 \sin2 \omega-3\pi\right]+3S_{0}n \right\} \cos I \label{eq:varomega} \\
  \left\langle\dot{\mathcal{M}}\right\rangle &=& 0. \label{eq:varM}
\end{eqnarray}


We point out that the variations of the semimajor axis, the eccentricity and the inclination are  pure effects of the time variability of the angular momentum, since they are proportional to $S_{1}$. On the other hand, in the variation of  the longitude of the ascending node (\ref{eq:varOmega}) and the argument of the periastron (\ref{eq:varomega}) there is a term proportional  to $S_{0}$, which is due to the usual (i.e. static) gravitomagnetic field.

From (\ref{eq:varOmega}) and (\ref{eq:varomega}) we can get the secular variation of the longitude of pericentre $\varpi=\Omega\cos I + \omega$:
\beq
\left\langle\dot \varpi \right\rangle = - 2\frac{G}{c^{2}a^{3}n} \left\{S_{1}\left[\frac 1 2 e \sin 2 \omega +2\pi\right]+2S_{0}n \right\} \cos I \label{eq:varpi}
\eeq

\section{Comparison with the observations} \label{sec:obs}

Here we will use (\ref{eq:vara})-(\ref{eq:varM}) to make contact with observations in real systems. Let us start from the system Earth-LAGEOS; indeed, the secular increase of the diurnal rotation period $P$ 
\begin{equation}\label{peri}
    \frac{\dot P}{P}=3\times 10^{-10}\ {\rm yr^{-1}}
\end{equation}
induces a linear decrease of the Earth's angular momentum  \cite{IERS}
\begin{equation}\label{anu}
    S = \alpha M R^2 \left(\frac{2\pi}{P}\right)= 5.86\times 10^{33}\ {\rm kg\ m^2 \ s^{-1}}
\end{equation}
of
\begin{equation}\label{varja}
    S_{1} = -5.6\times 10^{16}\ {\rm kg\ m^2\ s^{-2}}.
\end{equation}
For the LAGEOS satellite ($a = 12270$ km, $I=110$ deg), (\ref{varja}) and (\ref{eq:vara}) yield a decrease of its semimajor axis of only 
\begin{equation}\label{daaa}
    \left\langle\dot a\right\rangle = -2\times 10^{-14}\ {\rm m\ yr^{-1}},
\end{equation}
which is completely negligible with respect to the observed secular decrease of the semimajor axis of LAGEOS  of $-0.4$ m yr$^{-1}$ \cite{Parry}.

Another evaluation of the impact of the variation of these time-depending gravitomagnetic effects can be done by considering the double-pulsar system PSR J0737-3039
(\cite{burgay03,lyne04}). 

In order to give a rough estimate of the effect,  we  consider the gravitomagnetic field of PSR J0737-3039B, acting as a perturbation of the motion of the companion star PSR J0737-3039A. 
On setting  $S=\frac{2\pi}{P}\mathcal{I}$, where $\mathcal{I}$ is moment of inertia of  PSR J0737-3039B and $P$ its period, we may write
\begin{equation}\label{eq:psr1}
   \frac{\dot S}{S}=-\frac{\dot P}{P} 
\end{equation}
and, hence
\begin{equation}\label{eq:psr11}
   S_{1}=-2\pi \mathcal{I}  \frac{\dot P}{P^{2}} 
\end{equation}
In order to give a rough estimate, we can take for the moment inertia the value $\mathcal{I}=10^{38}$kg m$^{2}$ (see e.g. \cite{lorimer}); on considering the measured parameters of the system PSR J0737-3039 \cite{lyne04}, we obtain
\begin{equation}\label{eq:psr2}
    S_{1} = -7.8\times 10^{22}\ {\rm kg\ m^2\ s^{-2}}.
\end{equation}
On using (\ref{eq:vara}), where we take $\cos I=1$ to consider the most favorable condition, we get  a variation of the semimajor axis of 
\begin{equation}\label{eq:psr3}
    \left\langle\dot a\right\rangle = 4.7\times 10^{-11}\ {\rm m\ yr^{-1}},
\end{equation}
which is negligible, since the uncertainty on $a$ is about 8.1 $\times 10^{5}\ {\rm m}$ (see \cite{iorio08})

\section{Conclusions} \label{sec:conc}

We have studied the motion of a test particle in the gravitomagnetic field of a source whose spin angular momentum has a time-varying magnitude, starting from a recent solution of the GR field equations for a localized slowly rotating source in linear approximation. We have treated the gravitomagnetic acceleration as a perturbation of the Newtonian motion, and we have explicitly obtained the secular variations of the Keplerian  orbital elements due to this acceleration. Celestial bodies generally have variable angular momenta, so we have focused on some real systems, in order to evaluate the impact of these time-depending gravitomagnetic effects.  In particular, we have considered the LAGEOS satellites orbiting the Earth and the double pulsar system PSR J0737-3039. In both cases, we have evaluated the magnitude of the secular variation of the semimajor axis, and we have shown that the impact of the time variation of the spin angular momentum of the gravitational source is negligibly small, thus suggesting that it can be neglected in actual astronomical and astrophysical scenarios.

\section*{Acknowledgements}

We would like to thank  Dr. Luis Felipe Costa and Dr. Carlos Herdeiro for drawing our attention to their works and for their comments  on a previous version of this paper. MLR  thanks Prof. Guido Rizzi and Prof. Angelo Tartaglia for useful discussions.

\end{document}